\documentclass[aip,graphicx]{revtex4-1}
\usepackage{graphicx}

\begin{document}

\title{Engineering whispering gallery modes in MoSe$_2$/WS$_2$ double heterostructure nanocavities: Towards developing all-TMDC light sources}




\author{P. A. Alekseev}
\affiliation{Ioffe  Institute, St.~Petersburg, 194021, Russia}
\email{npoxep@gmail.com}
\author{I. A. Milekhin}
\affiliation{Ioffe  Institute, St.~Petersburg, 194021, Russia} 
\affiliation{Rzhanov Institute of Semiconductor Physics, 630090, Novosibirsk, Russia}

\author{K. A. Gasnikova}
\affiliation{Ioffe  Institute, St.~Petersburg, 194021, Russia}
\author{I. A. Eliseyev}
\affiliation{Ioffe  Institute, St.~Petersburg, 194021, Russia}
\author{V. Yu. Davydov}
\affiliation{Ioffe  Institute, St.~Petersburg, 194021, Russia}
\author{A. A. Bogdanov}
\affiliation{School of Physics and Engineering, ITMO University, St. Petersburg 197101, Russia}
\affiliation{Qingdao Innovation and Development Centre, Harbin Engineering University, Qingdao, Shandong 266000, China}
\author{V. Kravtsov}
\affiliation{School of Physics and Engineering, ITMO University, St. Petersburg 197101, Russia}
\author{A. O. Mikhin}
\affiliation{School of Physics and Engineering, ITMO University, St. Petersburg 197101, Russia}
\author{B. R. Borodin}
\affiliation{Ioffe  Institute, St.~Petersburg, 194021, Russia}
\author{A. G. Milekhin}
\affiliation{Ioffe  Institute, St.~Petersburg, 194021, Russia}
\affiliation{Rzhanov Institute of Semiconductor Physics, 630090, Novosibirsk, Russia}

\begin{abstract}

Transition metal dichalcogenides (TMDCs) have emerged as highly promising materials for nanophotonics and optoelectronics due to their exceptionally high refractive indices, strong excitonic photoluminescence (PL) in monolayer configurations, and the versatility to engineer van der Waals (vdW) heterostructures. In this work, we exploit the intense excitonic PL of a MoSe$_2$ monolayer combined with the high refractive index of bulk WS$_2$ to fabricate microdisk cavities with tunable light emission characteristics. These microdisks are created from a 50-nm-thick WS$_2$/MoSe$_2$/WS$_2$ double heterostructure using frictional mechanical scanning probe lithography. The resulting cavities achieve a 4-10-fold enhancement in excitonic PL from the MoSe$_2$ monolayer at wavelengths near 800 nm. The excitonic PL peak is modulated by sharp spectral features, which correspond to whispering gallery modes (WGMs) supported by the cavity. A microdisk with a diameter of 2.35 $\mu$m demonstrates WGMs with a quality factor of up to 700, significantly surpassing theoretical predictions and suggesting strong potential for lasing applications. The spectral positions of the WGMs can be finely tuned by adjusting the microdisk's diameter and thickness, as confirmed by theoretical calculations. This approach offers a novel route for developing ultra-compact, all-TMDC double heterostructure light sources with record-small size.

\end{abstract}

\maketitle %\maketitle must follow title, authors, abstract and \pacs

\section{Introduction}
Nanoscale light sources are in high demand for nanophotonics\cite{makarov2018nanoscale, zhou2015chip, wasisto2019beyond}, optical communication, and quantum computing\cite{o2007optical}. To enhance and control light intensity, polarization, and directivity, an optical cavity is used\cite{thompson2013coupling, maksimov2014circularly, alekseev2018half}. Several hybrid systems serving as a light-emitting media embedded into an optical cavity have been developed for lasers\cite{huang2017cspbbr3, najer2019gated}, light emitting diodes\cite{schubert1992resonant}, or complex light emitting devices\cite{yang2015laser}. However, achieving strong light confinement is challenging and requires high-refractive-index media. One of the most promising candidates for this purpose is transition metal dichalcogenides (TMDCs)\cite{eliseyev2023twisted}. In the visible and near-IR spectrum ranges, the refractive index of these materials can exceed 5 \cite{munkhbat2022optical}. A variety of exfoliable materials enables a wide tunability of target wavelengths from UV to near-IR \cite{wang20182d, tang2017electronic}. Additionally, their giant optical anisotropy provides additional ways to control light at the nanoscale \cite{ermolaev2021giant}. Recently, various photonic nanostructures utilizing bulk TMDC multilayers have been proposed and demonstrated \cite{ling2021all, li2024near, li2022tunable, munkhbat2020transition, zotev2023van, lee2023ultrathin, munkhbat2023nanostructured, zograf2024combining, ling2023deeply}. These papers have mainly focused on the waveguiding properties of the TMDC and light scattering properties by related nano-objects. 
However, TMDC materials can serve as effective light emitters. Over the last decade, it has been shown that monolayer TMDC materials having a direct bandgap \cite{mak2010atomically}, contrary to indirect bulk TMDCs, exhibit a strong excitonic photoluminescence (PL). This PL can be further enhanced by an external cavity resulting in the lasing from a monolayer (ML) TMDC \cite{wu2015monolayer}. Increasing the TMDC layer thickness reduces the PL intensity by several orders of magnitude due to the direct-indirect bandgap transition\cite{tonndorf2013photoluminescence}.

Recently, TMDC whispering gallery mode (WGM) resonators have drawn significant attention due to their strong light confinement in the out-of-plane direction and large Q-factors. For instance, Sung et al. demonstrated using immersive optics that 50-nm-thick WS$_2$ nanodisks, fabricated by electron-beam lithography, can support WGM resonances with a Q-factor of up to 400, which is sufficient for lasing \cite{sung2022room}. Our previous work proposed the chemical-free fabrication of WGM nanoresonators via a novel scanning probe lithography technique \cite{borodin2023indirect, borodin2021mechanical}. Using this technique, we demonstrated that 70-nm-thick MoSe$_2$ disks support WGM modes with a 100-fold photoluminescence enhancement and a pronounced Purcell effect. However, despite significant progress, the Q-factor of such cavities remains quite low compared to the best WGM cavities, which can demonstrate Q-factors on the order of 10$^6$ \cite{guha2017surface}. We believe that further increases in Q-factor in a single-material bulk TMDC cavity are severely limited by material losses. Bulk TMDCs are indirect bandgap semiconductors, and since absorption, enhancement, and emission occur in the same medium, the contribution of nonradiative band transitions is substantial. We believe that a double heterostructure approach of WGM lasers — i.e., a highly emissive quantum well within non-emitting cladding, might provide an additional degree of freedom in terms of emission wavelength and Q-factor control in all-TMDC optical resonators \cite{alferov2001nobel, wong2022iii}.

Here, we study all-TMDC whispering gallery mode heterostructure microdisks with varying diameters composed of a monolayer MoSe$_2$ core sandwiched between two WS$_2$ bulk cladding layers. The heterostructures, consisting of 25-nm WS$_2$/1 ML MoSe$_2$/25-nm WS$_2$, were fabricated using a standard dry-transfer protocol. Microdisks were then produced through a frictional mechanical scanning probe lithography technique, as described in Ref.~\cite{borodin2021mechanical}. The WS$_2$ cladding facilitates mode confinement, while the MoSe$_2$ monolayer acts as a quantum well, efficiently localizing excitons and enabling rapid radiative recombination. We characterized the microdisks through micro-photoluminescence spectroscopy and numerical modeling. The results reveal that the photoluminescence from the MoSe$_2$ monolayer is significantly modified by WGM-enhanced spectra, exhibiting linewidths as narrow as 1.4 nm and Q-factors reaching up to 700. The developed numerical model accurately reproduces the experimental observations and allows precise engineering of the microdisk parameters. This work establishes a novel, tunable platform of all-TMDC heterostructure WGM resonators with high Q-factors, offering exciting opportunities for advanced nanophotonic applications.

\section{Samples and Methods}
To create the WS$_2$/MoSe$_2$/WS$_2$ heterostructure, individual WS$_2$ and MoSe$_2$ flakes were obtained through mechanical exfoliation using the sticky tape method. The thickness of the exfoliated flakes was controlled by atomic force microscopy (AFM), while the excitonic properties of MoSe$_2$ monolayers were studied by micro-photoluminescence ($\mu$m-PL) spectroscopy. To observe the WGMs' optical resonances near the MoSe$_2$ excitonic emission energy, the cavity thickness should be at least 50 nm \cite{sung2022room}. Specifically, two WS$_2$ flakes, each 25 nm thick, were chosen for microresonator fabrication (see optical images in Fig. 1d). The thickness of the flakes was determined from AFM images presented in Supplementary material (Figure S1). The heterostructure was built with the layer sequence shown in Fig. 1a and 1d. To form an optical cavity from the vdW heterostructure, a scanning probe microscope Ntegra Aura (NT-MDT) was used for lithography. An AFM probe with a diamond tip (DRP-IN, Tipsnano) carved disks from the heterostructure. Since the TMDCs exhibit strong anisotropy in their in-plane and out-of-plane mechanical properties, the frictional mechanical scanning probe lithography (f-SPL) method was employed \cite{borodin2023indirect}. In this method, the probe was pressed against the structure with sufficient force to cut only several monomolecular vdW layers (Fig. 1b). By multiple repeating of this frictional process, a lithography pattern was created. Over 20 microdisks with various diameters ranging from 2 to 4.0 $\mu$m were fabricated. An optical image of several disks is presented in Fig. 1e.

\begin{figure}
\includegraphics[width=1\textwidth]{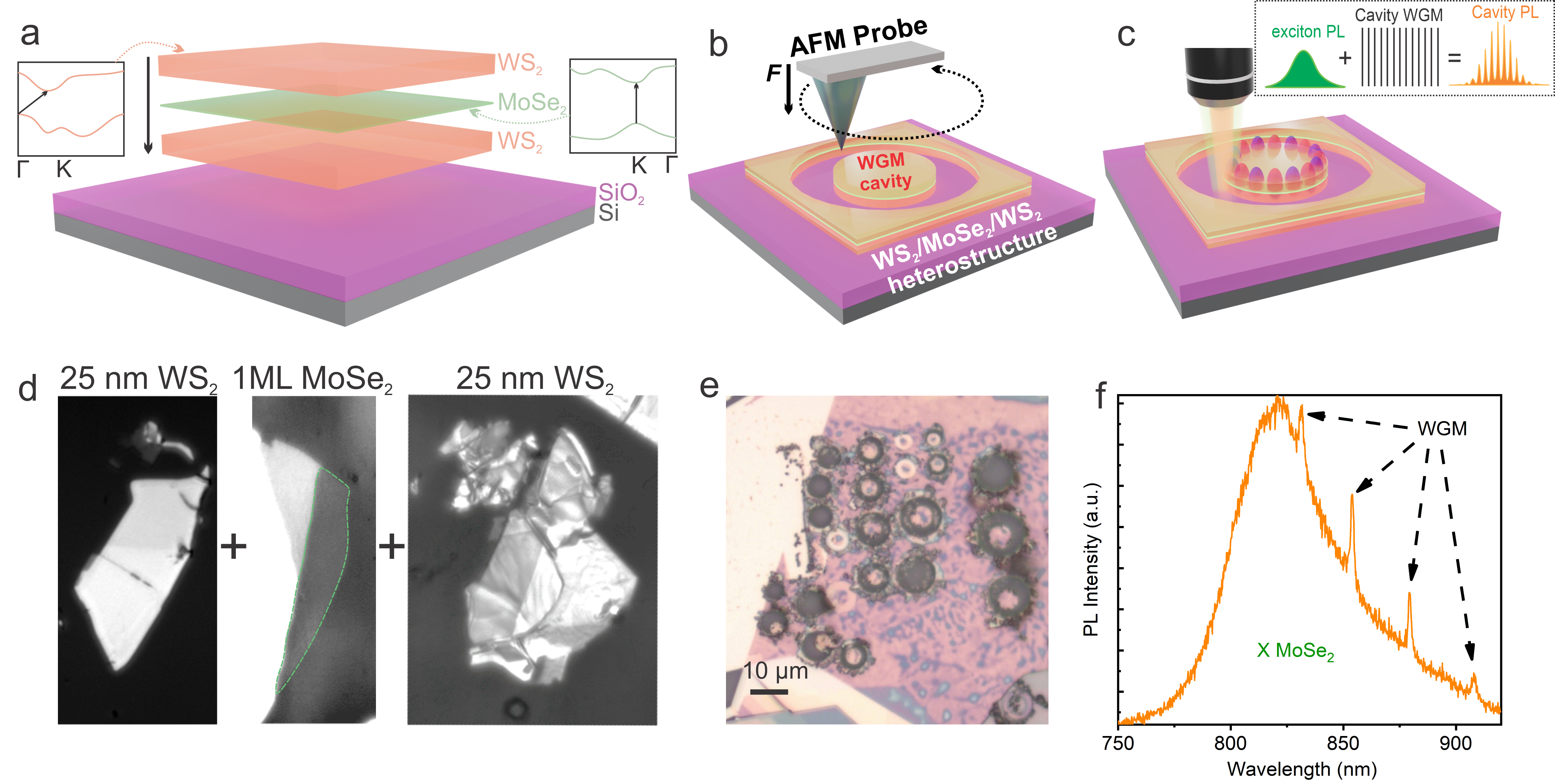}
\caption{\label{fig:1} WS$_2$/MoSe$_2$/WS$_2$ double heterostructure. a) Scheme of the layer sequence. Multilayer WS$_2$ is an indirect semiconductor, and monolayer MoSe$_2$ is a direct band gap semiconductor. Their bandgap structure is schematically shown in the insets. b) Scheme of the microdisk cavity carving by f-SPL. c) Geometry of the PL experiment and the sketch of PL spectra formation. d) Optical images of the WS$_2$ and MoSe$_2$ flakes before assembling vertical heterostructure. e) Optical image of the heterostructure with fabricated microdisk cavities. f) Typical experimental PL spectrum of a microdisk cavity.} 
\end{figure}

The fabricated microdisks were studied using using micro-PL spectroscopy. Measurements were performed with a Horiba XPlorA PLUS spectrometer at room temperature. The PL excitation and measurement were organized in a confocal scheme (Fig. 1c) using x100 objective with a numerical aperture (NA) 0.9, focusing the laser into a ~ 1 $\mu$m$^2$ spot. The disks were pumped with the 532 nm continuous wave laser at various pumping intensities (3 $\mu$W-3 mW). During the study, lateral mapping with a scanning step of 50-100 nm was also performed. 
\section{Results and discussion}
The spatial distribution of optical properties of the heterostructure with the fabricated microdisks was studied using PL spectra mapping (see Supplementary material Fig. S2). Disks with the sharpest (highest Q-factor) WGM lines were chosen and further mapped with a higher lateral resolution. Fig. 2a and 2b show the AFM image and topography profile for the disk with the highest Q-factor (700). In the figure, one can recognize the disk at the center, an outer area with removed TMDC material (darker ring), and the outermost area of the non-patterned TMDC heterostructure. The disk has a diameter of ~2.35 $\mu$m and a thickness of ~50 nm. The roughness of the disk surface of about 5-10 nm is due to residual polymeric glue and air gaps remaining on the surface and interface after the formation of the heterostructure. The outer area of the disk has some topographical features that can be attributed to ripped TMDC material. The main part of the removed material is gathered on the surface of the outermost area of the image.

Figure 2c shows a spatial map of PL intensity integrated over the 735-925 nm spectral range, obtained with a pumping intensity of 320 $\mu$W. As one can see from the figure, the highest PL intensity is observed at the center of the disk and gradually decreases towards the disk periphery. This can be explained by the nonradiative recombination at the disk edges\cite{borodin2021photoluminescence} and partial overlap of the excitation spot and the disk area at the periphery. Interestingly, the centers of the disks exhibit an increase in PL intensity by a factor of 4-10 compared with the clean, non-patterned heterostructure (See Supplementary material Fig. S2).
PL spectra obtained from various regions of the microdisk are presented in Fig. 2d, with corresponding regions marked in Fig. 2c (points 1-5). The spectra exhibit a broad peak with a maximum at 810-820 nm, which can be attributed to excitonic emission (A-exciton) of the MoSe$_2$ monolayer\cite{tonndorf2013photoluminescence}. Conventionally, PL spectra of a MoSe$_2$ ML on the SiO$_2$ have a maximum at 790-800 nm, and the shift of the PL maximum in the heterostructure is due to the dielectric screening caused by the WS$_2$ cladding\cite{raja2017coulomb}. 

\begin{figure}
\includegraphics[width=1\textwidth]{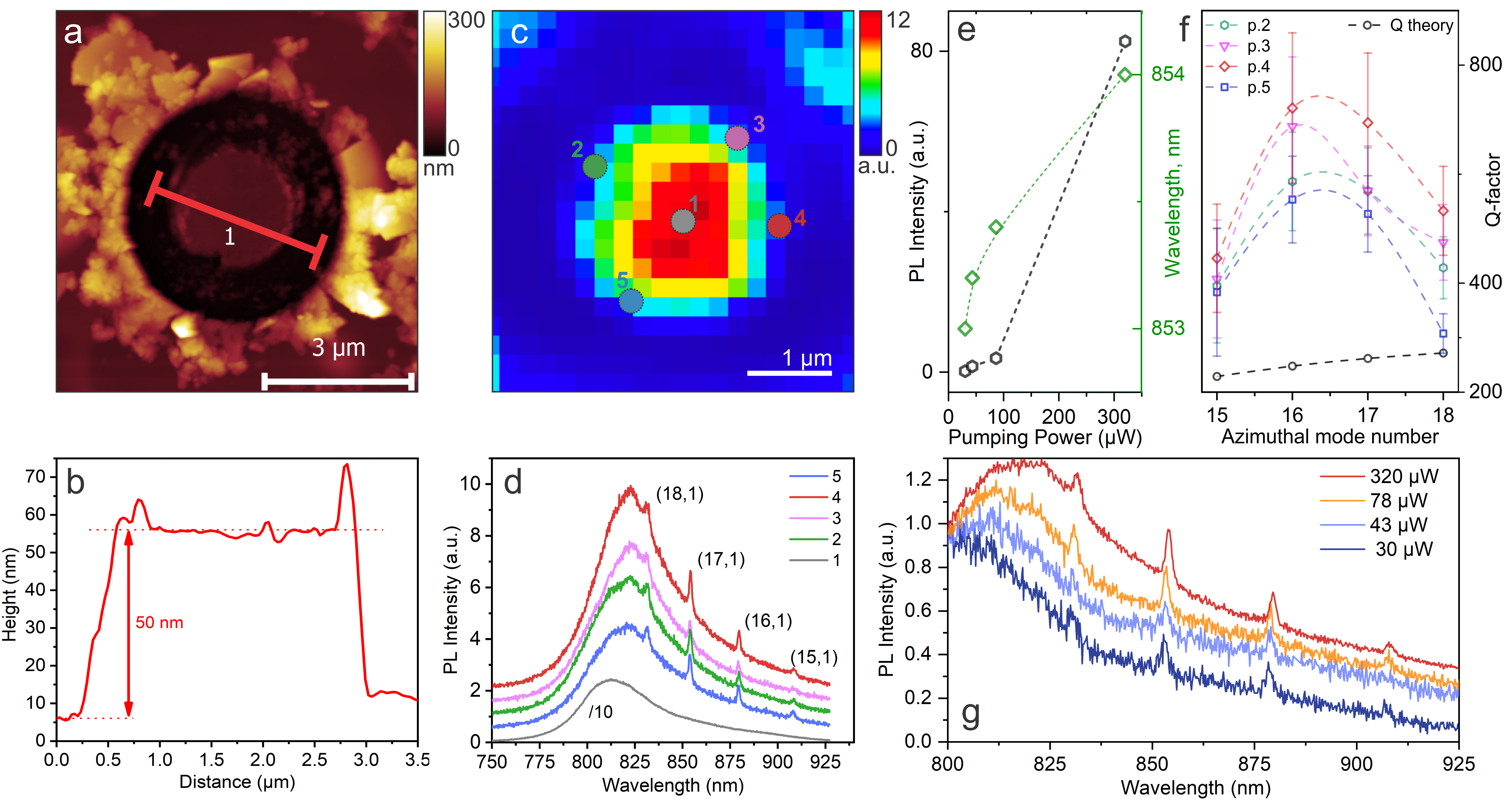}
\caption{Microdisk cavity with a 2.35 $\mu$m diameter and 50 nm thickness. a) AFM image and corresponding profile (b). c) PL intensity map and (d) PL spectra obtained at corresponding points 1-5. e) PL intensity (black squares) and position of the WGM peak at 853 nm (green circles) measured with various pumping intensities. f) Experimental and calculated (open circles) Q-factors of the WGM narrow peaks in Fig. 2d. g) PL spectra obtained at various pumping intensities.} 
\end{figure}

The broad excitonic peak in the spectra 2-5 is modulated by narrow features at 831, 854, 879, and 909 nm. These features can be assigned to the WGM modes. To classify the WGM modes, their azimuthal and radial numbers were determined from calculations (Fig. 3) and labeled in Fig. 2d. 
WGM spectra in WS$_2$ disks on SiO$_2$ substrates were simulated using COMSOL Multiphysics (version 6.1). As the pattern has rotational symmetry, we used a '2D Axisymmetric' model to increase the computation speed. The model uses 'Electromagnetic Waves, Frequency Domain' physics from the optics module. The monolayer of MoSe$_2$ was neglected in the calculations. The material parameters of WS$_2$, SiO$_2$, and Si were taken at a wavelength of 855 nm\cite{vyshnevyy2023van}. Due to the high optical birefringence, the in-plane refractive index (n$_o$) is 4.13, while the out-of-plane refractive index (n$_e$) is 2.52. The extinction coefficient k=0.0077.    Radiative losses are accounted for via perfect match layers surrounding the disk. The electromagnetic field was visualized in 3D space by revolving the 2D domain, accounting for the azimuthal number of WGMs.

The positions of the narrow peaks (Fig. 2d) remain the same for points 2-5. Therefore, the excitation conditions do not perturb the mode structure of the cavity. Note that no narrow peaks were observed in the spectral range of 750-810 nm (Fig. 2d) due to the increasing WS$_2$ light absorption ($k$=0.02 at 810 nm). For the same reason, only WGMs with a unity radial number propagating near the very edge of the disk were experimentally observed. Additionally, the light absorption of MoSe$_2$ resonantly grows, and the extinction coefficient at the A-exciton maximum is 0.7.\cite{hsu2019thickness}
Figure 2f shows the quality factor (Q-factor) obtained for the various WGMs from spectra 2-5 (Fig. 2d), along with calculated values. A non-monotonic dependence of the experimentally obtained Q-factor on wavelength was observed. A relatively low value of Q-factor of the modes with high azimuthal numbers (observed in PL spectra at the short wavelengths) is due to the WS$_2$ absorption (not taken into account in the calculations)\cite{vyshnevyy2023van}, while the modes with low azimuthal numbers (seen at the long wavelengths) are not well-confined in the cavity due to the small thickness and diameter of the cavity (see the Q-factor values in Fig. 3). Interestingly, the experimentally obtained Q-factors reached 700, which are higher than the theoretically predicted values.
\begin{figure}
\includegraphics[width=1\textwidth]{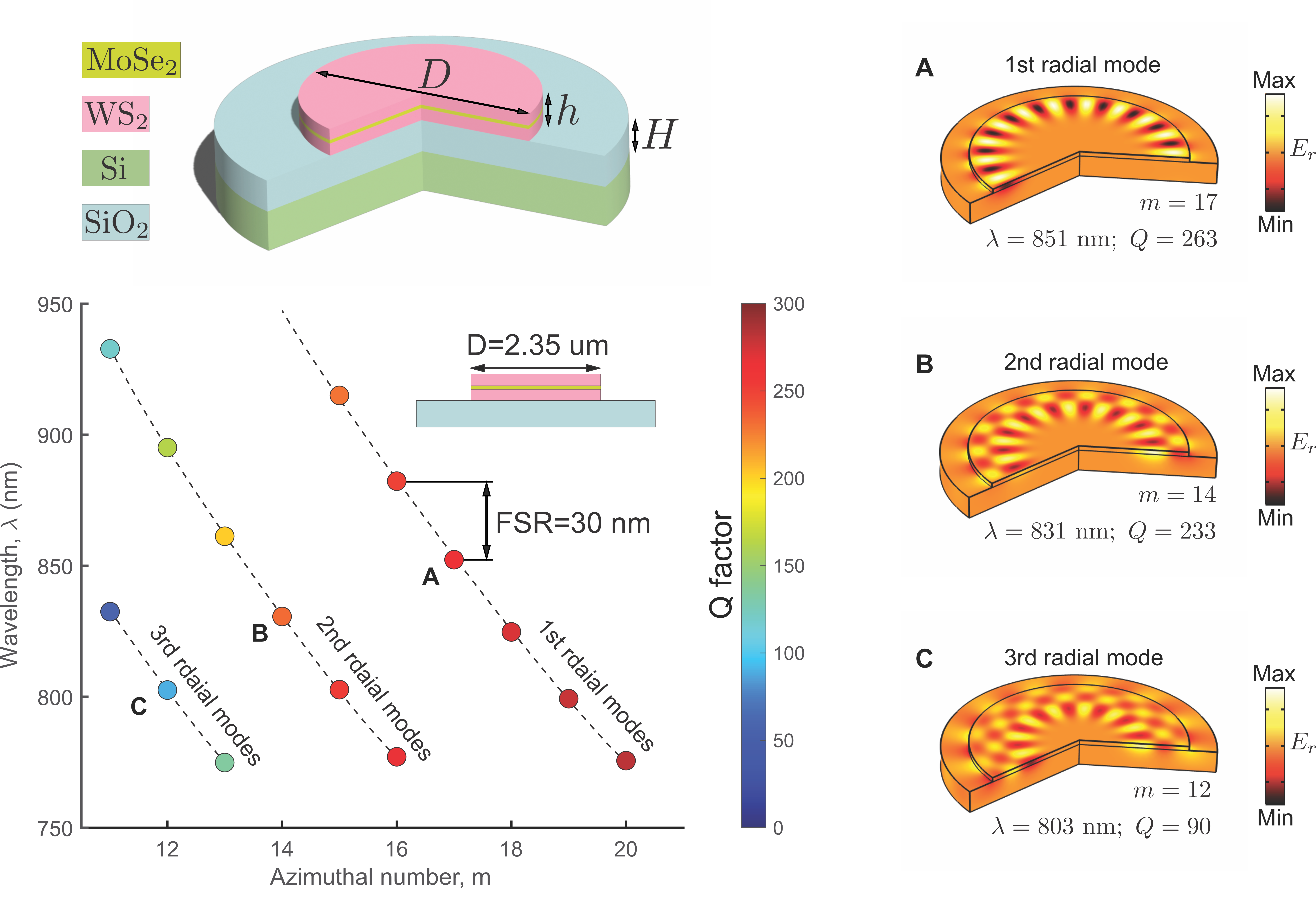}
\caption{\label{fig:3} Calculation of the WGM spectrum and Q factors for the microdisk cavity with a 2.35 $\mu$m diameter and 50 nm thickness. Scheme of the structure used in the calculations (left upper inset). Electrical field distribution for the 1st (A), 2nd (B), and 3rd (C) radial modes.} 
\end{figure}
To further understand the light-emitting properties of the disk cavity, the PL spectra mapping was performed for various pumping intensities. The PL spectra obtained at point 2 (Fig. 2c) with pumping in a 30-320 $\mu$W power range are presented in Fig. 2g. The spectral intensity is normalized to unity, and the spectra are vertically shifted for clarity. Figure 2e shows non-normalized PL intensity for the various pumping intensities. It is worth noting that the PL intensity grows non-linearly with pumping. At first glance, this dependence shows a threshold, which presumably indicates laser generation. However, the dependence is the same for all points (see 1-5 in Fig. 2d) in the cavity. Narrow WGM peaks grow at the same rate as the broad excitonic peak. This behavior can be explained by the saturation of the absorption channel of the WS$_2$ with increased pumping. The line widths of the narrow WGM peaks were also analyzed for various pumping intensities, and no changes in the widths were observed within the measurement error (noise level is too high for the 30 and 43 $\mu$W pumping).

Increasing the pumping leads to the redshift of the WGM peaks. The same behavior was previously observed for the WS$_2$ microdisks and was associated with the heating of the structure, which causes changes in structural parameters due to thermal expansion and alteration of optical parameters with increasing temperature \cite{sung2022room}. The shift of the peak intensity for the WGM at 853 nm is shown in Fig. 2e (green circles). Further increase of the pumping beyond 320 $\mu$W damages the disk, and the PL quenches.

Figure 3 shows the calculation results for the disk with a diameter of 2.35 $\mu$m. The obtained free spectral range (FSR) value agrees well with that experimentally observed at 855 nm. At shorter wavelengths, the experimental FSR decreases faster than the calculated value due to the increasing refractive index, which is not accounted for in the model. From the figure, it follows that the WGM with a higher azimuthal number has a higher Q-factor. Additionally, the Q factors of the WGM of the first and second radial numbers are close to each other.

To increase the Q-factor, the cavities with a larger diameter were created. Fig. 4 shows PL spectra obtained for disks with larger diameters and thicknesses (Fig. 4c). As in the case of the spectra in Fig. 2, a broad excitonic peak modulated by the narrow WGM peaks is observed for all diameters. Calculated WGM numbers are also labeled in the spectra, and detailed calculation results (Q-factors) are presented in the Supplementary material (Fig. S3-S5). Interestingly, the increase in the diameter does not lead to the narrowing of WGM features. Figure 4e shows experimental and calculated Q-factors obtained for disks and their narrow peaks in the 840-860 nm spectral range. The 2.35 $\mu$m diameter disk in Fig. 4b (the same disk as in Fig. 2) shows a significantly higher Q-factor than disks with larger sizes.
\begin{figure}
\includegraphics[width=1\textwidth]{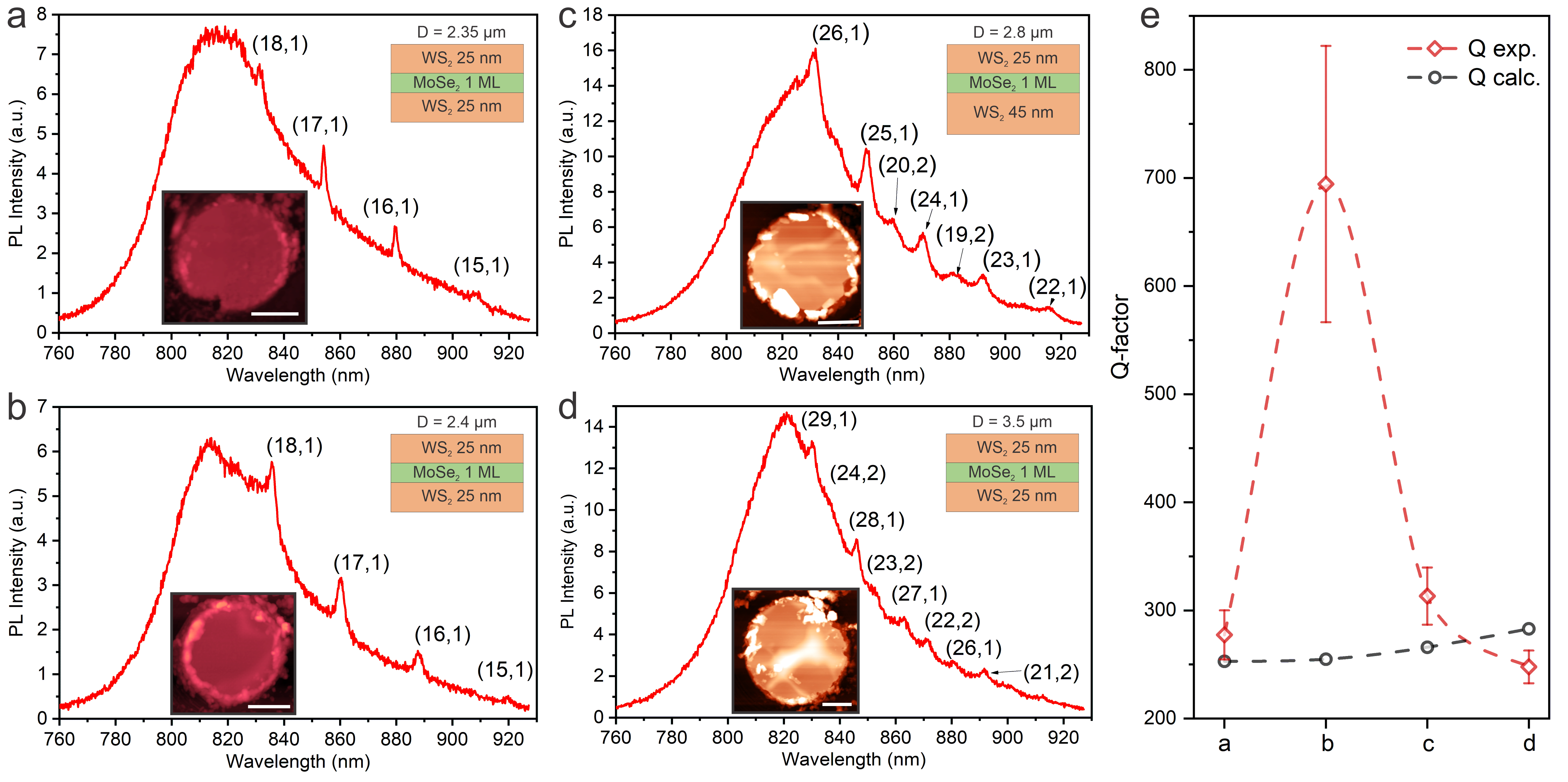}
\caption{\label{fig:4} PL spectra for various microdisk cavity sizes a)-d). Right insets represent diameter and layer thickness. Central insets show AFM images of the corresponding cavity. The scale bar is 1 $\mu$m. e) Experimental (black squares) and calculated Q-factors of the WGM peaks in the 840-860 spectral range.} 
\end{figure}

Moreover, the larger disks exhibit experimental Q-factors substantially lower than the calculated values, while the smallest disk has a Q-factor higher than the calculated value. Probably the smallest disk with the narrowest peaks works in an amplified spontaneous emission mode\cite{sung2022room}. It is worth noting that several disks with a diameter smaller than 2.35 $\mu$m were created, which showed Q-factors below 200 (not shown here).
To understand the low Q-factor values for the larger disk, the AFM images in Fig. 4 were analyzed. All the disks have significant roughness on their edges. This roughness, on the one hand, should allow the coupling of a pumping beam with the cavity and the detection of the WGM in the PL experiment. On the other hand, the roughness decreases the Q factor of the cavity. Increasing the disk diameter increases the impact of the roughness on the Q factor due to the localization of the mode at the very edge of the disk.
Spectra in Fig. 4c and 4d are modulated by WGMs with the first and second radial numbers. Calculated Q factors (see Supporting information Fig. S3-S5) are nearly the same for the first and second radial numbers, however, the PL intensity of the second modes is lower. This is presumably due to lesser light outcoupling of the second radial mode in the PL experiment.

Thus, increasing the microdisk's diameter does not increase the Q-factor of the WGM due to the high roughness of the disk edges. A previous study, where disk cavities were formed from the bulk MoSe$_2$ by frictional SPL, presented cavities with smooth edges\cite{borodin2023indirect}. The rough edges of the heterostructure disks presented here are due to the thin air and polymer gaps at the heterolayer interfaces. These randomly distributed gaps perturb the shear modulus of the heterostructure, and the cutting force is not well-controlled during the f-SPL processing. This is the reason for the higher roughness of the heterostructure disks. Further improvement of the f-SPL of the TMDC heterostructures can be performed by scanning probe ironing of the heterostructure before the f-SPL\cite{borodin2023effect, rosenberger2018nano}. Nevertheless, conventional lithography methods, including the etching step, also do not provide perspective for the heterostructures with the gaps at the interface. To date, only homostructural multilayer TMDC disk cavities have been demonstrated. For example, WS$_2$ disks with smooth edges formed by conventional lithography had a Q-factor of 400\cite{sung2022room}, which is comparable with the results shown here. An additional advantage of the f-SPL is that the formed disk cavity can be mechanically manipulated and positioned at the desired place (see Supplementary material S6).
The Q-factor of the all-TMDC double heterostructure cavities can also be increased by choosing cladding with a wider bandgap. Indeed, choosing SnS$_2$ instead of the WS$_2$ allows creating the cavities with near-zero absorption at the wavelengths of the exciton in MoSe$_2$ and WS$_2$ monolayers. For infrared (IR) applications, a pair of WS$_2$/MoTe$_2$/WS$_2$ looks promising, due to near-IR luminescence of the MoTe$_2$ ML.

\section{Conclusions}

To conclude, we present a novel all-TMDC double heterostructure comprising two WS$_2$ bulk layers and a MoSe$_2$ monolayer with strong excitonic photoluminescence. The double heterostructure was fabricated using standard exfoliation and transfer techniques. High-quality microdisk cavities, with diameters ranging from 2 to 4 $\mu$m, were precisely carved from the heterostructure using frictional scanning probe lithography. These microdisks exhibit a 4--10-fold enhancement in the excitonic photoluminescence of the MoSe$_2$ monolayer compared to the pristine heterostructure. The broad excitonic PL peak is modulated by narrow spectral features corresponding to whispering gallery modes (WGMs). The spectral positions of the WGMs are finely tuned by adjusting the disk thickness and diameter. The observed WGM PL linewidth is less than 1.4 nm, and the Q-factor reaches 700 for a microdisk with a diameter of 2.35 $\mu$m and a thickness of 50 nm, marking a significant step toward lasing nanostructures. These results highlight a promising approach for fabricating all-TMDC heterostructure nanophotonic light-emitting devices, offering precise wavelength tunability and an ultra-compact form factor.

The authors have no conflicts to disclose.

\begin{acknowledgements}

The work is supported by Russian Science Foundation, 24-12-00209, https://rscf.ru/project/24-12-00209/.

\end{acknowledgements}
\section*{Data Availability Statement}
The data that support the findings of this study are available from the corresponding author upon reasonable request.

\bibliography{Bibl}

\end{document}